\documentclass[prl, preprintnumbers ,showpacs,amsmath,amssymb, superscriptaddress,twocolumn]{revtex4}
\usepackage{bm}
\usepackage{amsmath}
\usepackage{amssymb}
\usepackage{amsthm}
\usepackage{amsfonts}
\usepackage{enumerate}
\usepackage{latexsym}
\usepackage{times}
\usepackage{graphicx}
\usepackage{color}
\usepackage{makeidx}
\usepackage{ifpdf}

\begin{document}

\title{Quantum confinement of Mott electrons in ultrathin LaNiO$_{3}$/LaAlO$_{3}$ superlattices}

\author{Jian~Liu} \email{jxl026@uark.edu}
\affiliation{Department of Physics, University of Arkansas, Fayetteville, Arkansas 72701}
\author{S.~Okamoto}
\affiliation{Materials Science and Technology Division, Oak Ridge National Laboratory, Oak Ridge, Tennessee 37831}
\author{M.~van~Veenendaal}
\affiliation{Advanced Photon Source, Argonne National Laboratory, Argonne, Illinois 60439}
\affiliation{Department of Physics, Northern Illinois University, DeKalb, Illinois 60115}
\author{M.~Kareev}
\affiliation{Department of Physics, University of Arkansas, Fayetteville, Arkansas 72701}
\author{B.~Gray}
\affiliation{Department of Physics, University of Arkansas, Fayetteville, Arkansas 72701}
\author{P.~Ryan}
\affiliation{Advanced Photon Source, Argonne National Laboratory, Argonne, Illinois 60439}
\author{J.~W.~Freeland}
\affiliation{Advanced Photon Source, Argonne National Laboratory, Argonne, Illinois 60439}
\author{J.~Chakhalian}
\affiliation{Department of Physics, University of Arkansas, Fayetteville, Arkansas 72701}

\date{\today}

%%% BEGIN DOCUMENT
\begin{abstract}
  We investigate the electronic reconstruction in (LaNiO$_{3}$)$_n$/(LaAlO$_{3}$)$_3$ ($n=$3, 5 and 10) superlattices due to the quantum confinement (QC)  by d.c. transport and resonant soft x-ray absorption spectroscopy. In proximity to the QC  limit, a Mott-type transition from an itinerant electron behavior to a localized state is observed. The system exhibits tendency towards charge-order during the transition. $ab$ $initio$ cluster calculations are in good agreement with the absorption spectra, indicating that the apical ligand hole density is highly suppressed resulting in a strong modification of the electronic structure. At the dimensional crossover cellular dynamical-mean-field calculations support the emergence of a Mott insulator ground state in the heterostructured ultra-thin slab of LaNiO$_{3}$.
\end{abstract}

\maketitle
%\section{Introduction}

Prompted by the discovery of high-$T_{\rm c}$ superconductivity in cuprate compounds there has been a surge of activity to discover materials with  even higher transition temperature \cite{Bednorz}. Recent  remarkable advances in synthesis of artificial layers of complex oxides along with the progress in computational methods have re-energized the search for novel superconductors outside of the cuprate family \cite{Reyren,Takahashi}.

Towards the  challenge,  a recent proposal  has been put forward  to use heterostructuring and orbital engineering  to turn hole-doped alternating  unit-cell thin layers  of  a correlated metal LaNiO$_3$ (LNO) and band-gap dielectric LaAlO$_3$ (LAO) \cite{Chaloupka}. The proposal utilizes the Ni$^{III}$ $3d^7$ low-spin state in the bulk with a single unpaired electron occupying the degenerate ($d_{x^2-y^2}/d_{3z^2-r^2}$) $e_g$ orbital whose nodes point to the planar and apical ligands of the octahedra, respectively. Both \textit{t}--\textit{J} and LDA+DMFT calculations \cite{Chaloupka,Hansmann} suggest that the quantum confinement together with the electronic correlations should make it possible to localize or empty the $d_{3z^{2}-r^2}$ band leaving the conduction electron in the $d_{x^{2}-y^{2}}$ band. Epitaxial strain is also suggested as a mean of the orbital control to manipulate the $d_{3z^2-r^2}$ orbital to appear above $d_{x^{2}-y^{2}}$ and  play the analogous role of the axial orbital of the high-$T_{\rm c}$ cuprates.

Despite the attractive simplicity of the structure,  the experimental realization of LNO/LAO superlattice (SL) presents  a large degree of ambiguity in  selecting a specific electronic ground state caused by  (i) the intrinsic propensity  of ortho-nickelates to charge-ordering \cite{Mazin,Catalan0}, (ii) the orbital polarization due to chemical confinement imposed by interfacial bonding  \cite{Han,Freeland},  (iii) polar discontinuity \cite{Nakagawa,Liu},  (iv) epitaxial constraint on the crystal symmetry \cite{Chakhalian2,May} and (v) quantum confinement \cite{Seo,Hotta}. Even for the undoped LNO/LAO SL (i.e. the proposed parent cuprate-like compound), the synergetic action of these phenomena will likely  render  the theoretical problem intractable to \textit{a priori} predict whether the cuprate-like physics can be experimentally realized.

Here we report on emergence of a Mott-type metal-insulator transition (MIT) at the dimensional crossover in the experimentally  realized  (LNO)$_n$/(LAO)$_3$ SLs. By using transport measurement and soft x-ray absorption spectroscopy in combination with cluster calculations and cellular dynamical-mean-field theory calculations, we  demonstrate the presence of a charge-localized ground state at  the  quantum confinement limit caused by strongly enhanced correlations and broken translational symmetry across the interface. In  sharp contrast to conventional bandwidth-control, the observed MIT is a continuous one with a critical region due to an emerging correlated gap imposed by confined dimensionality. The system also exhibits tendency towards charge-ordering as a competing state during the transition. Microscopically, the crossover  from the localized to itinerant  behaviors is driven by a strong competition between  the reduced covalency of the Ni-O-Al bond and the recovery of the ligand hole density within the LNO slab.
\\

%\section{Experimental}
%\section{Growth}

High-quality  epitaxial [(LNO)$_n$/(LAO)$_3$]$_N$ SLs ($n=3$, 5 and $10$ with $N=10$, 4 and 2 respectively) were grown by  laser molecular beam epitaxy with \textit{in situ} monitoring by Reflection High Energy Electron Diffraction, as described in Ref.\cite{Kareev1,Liu}; +2.2\% tensile strain is applied with TiO$_{2}$-terminated (001) SrTiO$_3$ substrates \cite{Kareev2}. Atomic force microscopy imaging reveals that the substrate surface morphology is well maintained after the growth; Symmetric scans and reciprocal space mapping by synchrotron-based x-ray diffraction indicate coherently grown single crystal heterostructures with the fully strained state\cite{Kareev1}.

%\section{Transport}
\begin{figure}[t]\vspace{-0pt}
\includegraphics[width=8.5cm]{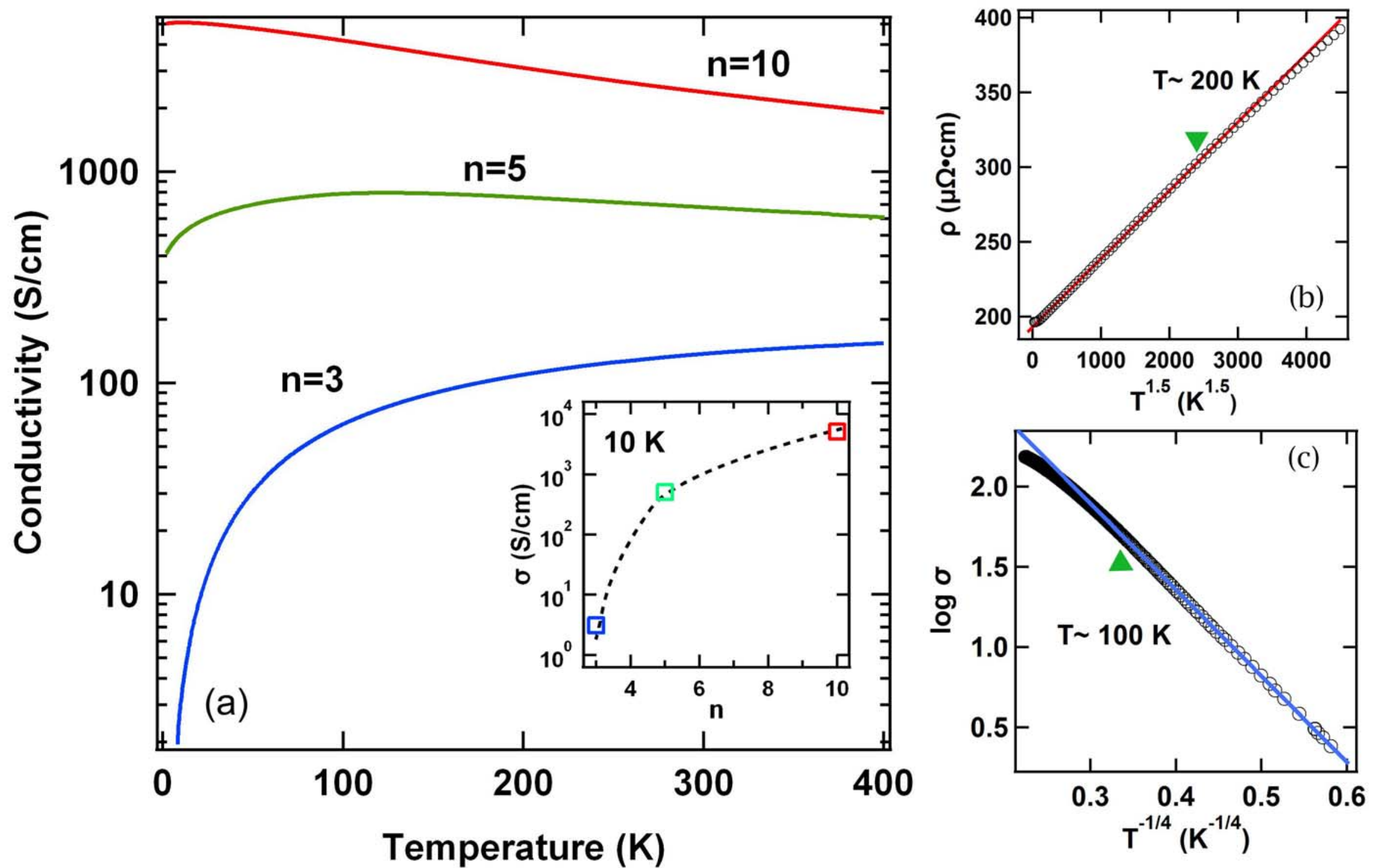}
\caption{\label{transport}  (a) Conductivity of (LNO)$_n$/(LAO)$_3$ SLs versus temperature. Inset: conductivity versus $n$ at 10 {\rm K}. The dashed line is a guide for eye. (b) Resistivity as a function of $T^{1.5}$ at $n=10$. The red solid line is a guid for eye. (c) $d\ln\sigma$ as a function of $1/T^{1/4}$ for the $n=3$ SL shows the low temperature three dimensional variable range hopping behavior.}
\end{figure}

To explicate the evolution of the ground state due to quantum confinement, d.c. transport measurements from 400 {\rm K} to 2 {\rm K} were performed in the $van$ $der$ $Pauw$ geometry with a commercial physical properties measurement system (PPMS, Quantum Design). The obtained temperature dependent conductivity, $\sigma(T)$, of the SL series is shown in Fig.~\ref{transport}(a). A transition from metallic to insulating behavior can be clearly seen with decreasing $n$. The inset of Fig.~\ref{transport}(a) also demonstrates orders of magnitude change in $\sigma$ across the series taken at 10 {\rm K}. As seen, in analogy to the bulk LNO, the $n=10$ SL exhibits a highly metallic behavior ascribed  to the three dimensional regime. On the other hand, unlike the bulk which is known to follow a $T^{2}$ Fermi liquid behavior at low temperatures \cite{Zhou}, the low-temperature resistivity $\rho$ at $n=10$ shows a peculiar $\propto$ $T^{3/2}$ dependence below 200 {\rm K} (see Fig.~\ref{transport}(b)). This power law dependence has been attributed to scattering by bond-length fluctuations \cite{Rivadulla} and indicates the proximity to a Mott-type metal-to-insulator transition (MIT) from the itinerant side due to enhanced electronic correlations in the quantum confinement regime.

The intermediate thickness $n=5$ SL still shows a clear metallic behavior at high temperatures but reaches a conductivity maximum at $T_{\rm max}$= 120 {\rm K}. While the metallicity of LNO films under large tensile strain is known to cease when the thickness  is $\leq3 {\rm nm}$ \cite{Scherwitzl}, it is remarkable that the high temperature metallic behavior is well maintained in such a thin slab ($<2 {\rm nm}$). This result unambiguously  highlights  the exquisite role of the confinement structure with interfaces in defining the transport  properties of SLs as compared to the bulk LNO and single layer films. Notice, the conductivity $\sigma$ here is also well below the Ioffe-Regel limit \cite{Mott} ($\sim2000$ {\rm S/cm}), manifesting the modification of the correlation nature of the observed behaviors and the minor role of disorder. Below $T_{\rm max}$, $\sigma(T_{\rm max})-\sigma(2~{\rm K})$ ($\sim$381.6 S/cm) is as large as $\sigma(2~{\rm K})$ and almost twice of $\sigma(T_{\rm max})-\sigma(400~{\rm K})$. Neither hopping conductivity nor quantum corrections is found to be able to account for this $d\sigma/dT>0$ region.
This transitional behavior explicitly indicates that all the involved interactions and relevant energy scales, like electron correlations  and the bandwidth, are of a similar strength and strongly competing with each other. Thus, $n=5$ falls into a critical region of the confinement-controlled MIT. Upon further decreasing $n$ to 3, $d\sigma/dT>0$ at all temperatures and $\sigma(T)$ is found to follow a variable range hopping behavior below 100 {\rm K} (see Fig.~\ref{transport}(c)), indicating that the system enters the insulating phase at $n=3$. The evolution with $n$ shown above points to an opening gap in the excitation spectrum.

%\section{XAS}

\begin{figure}[t]\vspace{-0pt}
\includegraphics[width=8.5cm]{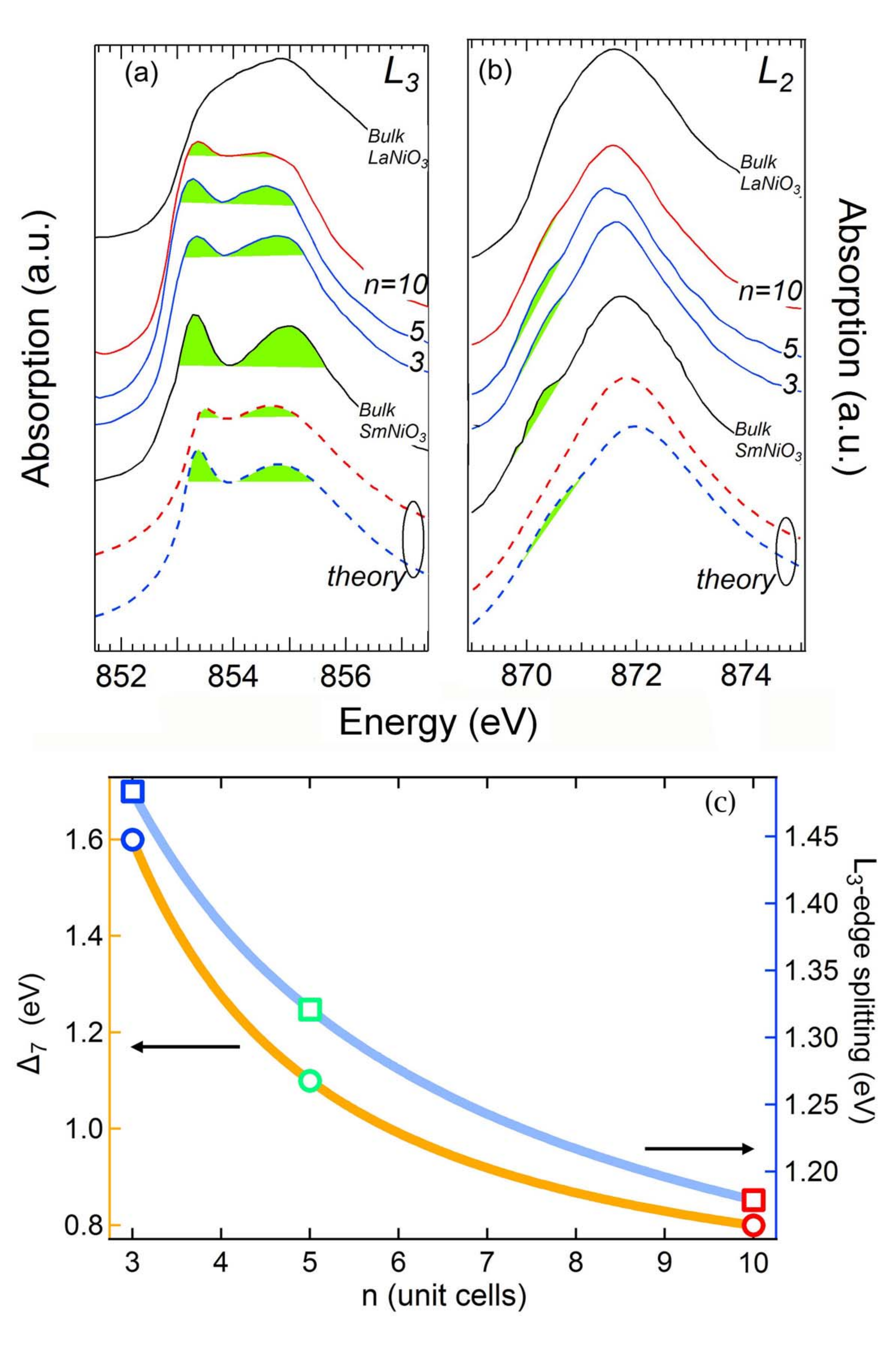}
\caption{\label{LNO}
X-ray absorption spectra of (LNO)$_n$/(LAO)$_3$ SLs at the $L_3$-edge (a) and $L_2$-edge (b) for $n=3,5$, and 10 together with the theoretical calculations done for a NiO$_6$ cluster for a charge transfer energy of 0.8 (red) and 1.6 (blue) eV. Spectra of bulk LNO and SmNiO$_3$ \cite{Freeland2} are included for comparison. The shaded (green) areas indicate the most significant changes during the evolution. (c) The observed multiplet splitting energy (square) and the corresponding calculated charge transfer energy (circle) versus $n$. Solid lines are fits to power-law.
}
\end{figure}

To gain microscopic insight into the effect of confinement on the nature of the MIT, we carried out detailed soft x-ray absorption measurements. Since resonant x-ray absorption is local and element specific probe, it is well suited for investigating electronic structure of  buried layers. All spectra were acquired in the bulk-sensitive fluorescence yield  mode at the Ni $L$-edge at the 4ID-C beamline of the Advanced Photon Source \cite{footnote1}. Precise information on the Ni charge state was obtained by aligning all the spectra to a NiO standard which was measured simultaneously with the SLs. The energy  resolution was set  at  100 meV.
The normalized spectra obtained at 300 K are shown in Fig.~\ref{LNO}(a,b). For direct comparison, the $L$-edge spectra of bulk LNO and SmNiO$_3$ \cite{Freeland2} are included. As seen, the energy position and the lineshape of the SLs are close to those of the bulk RENiO$_3$ (RE = rare earth) \cite{Medarde, Piamonteze}, indicative of a Ni$^{III}$ charge state of the $3d^7$ low-spin configuration.  On the other hand, while the multiplet effects are smeared in the metallic bulk LNO, the $L_3$-edge spectra of the SL show a pronounced two-peak structure (shaded in green), closely  resembling the absorption spectrum of the bulk SmNiO$_3$ (see Fig. 2.(a)), which in the bulk is a well studied case for the charge-ordered (2Ni$^{III}$$\rightarrow$Ni$^{II}$+Ni$^{IV}$) insulating state \cite{Mazin,Catalan0}. As $n$ decreases this multiplet effect becomes more pronounced, particularly in moving from $n=10$ to $5$. The energy separation of the two split peaks is also clearly increasing as shown in Fig.~\ref{LNO}(c). The corresponding trend at the $L_2$-edge emerges as a developing low-energy shoulder at about 870.5 eV (see Fig. 2.(b)). The observed evolution is reminiscent of that reported for the bulk RENiO$_3$ when crossing the metal-insulator boundary \cite{Piamonteze}, unambiguously revealing the carrier localization in the presence of a developing correlated gap in the quantum confined LNO slabs and a latent tendency to charge ordering during the three-to-two dimensional crossover. In accordance with the observations from d.c. transport, the emergence of the multiplet even in the relatively thick LNO slab ($n=10$) lends solid support to  strongly confinement-enhanced correlations. Meanwhile, when subject to this enhancement, the inclination to charge ordering corroborates its role as a primary competing ground state not only in the bulk \cite{Mazin} but also in the predicted heterostructured LNO \cite{Chaloupka}.

To shed additional light on the observed spectroscopic results, we performed $ab$ $initio$ cluster calculations on a NiO$_6$ cluster. The model Hamiltonian includes on-site energies, crystal field, Coulomb interaction including the full multiplet structure and spin-orbit coupling \cite{Veenendaal}. The majority of the parameters are equivalent to those in Ref.\cite{Veenendaal}. The monopole part of the Coulomb interaction is set at 6 eV. The major configurations used in the calculation are $\underline{d}_{e_g\uparrow}^2\underline{d}_{e_g\downarrow}$ and $\underline{d}_{e_g\uparrow}^2\underline{L}_{e_g\downarrow},$ where $\underline{d}_{e_g\sigma}$ and $\underline{L}_{e_g\downarrow}$ stand respectively for a $3d$ hole and a ligand hole with $e_g$ symmetry and spin $\sigma=\uparrow,\downarrow$. Note that trivalent nickel compounds, such as LNO, are significantly more covalent than divalent systems as the charge-transfer energy $\Delta_n=E(d^{n+1}{\underline L})-E(d^n)$ decreases due to the change in electron count with $\Delta_7\cong \Delta_8-U$, leading to strong mixing between the $d^8{\underline L}$ and $d^7$ configurations. The degree of Ni-O covalency here is varied by changing the charge-transfer energy $\Delta_7$. Figure~\ref{LNO}(a) shows the result of calculations when $\Delta_7$ was varied from 0.8 eV to 1.6 eV. The calculation strongly implies that when the local Ni-O covalency is reduced with increasing $\Delta_7$, the observed evolution of the $L_3$-edge multiplet splitting with $n$ is well reproduced. To quantify the $n$-dependence, the value of $\Delta_7$ for each SL is obtained by matching the corresponding size of the splitting. As can be seen in Fig.~\ref{LNO}(c), the $n$-dependence of $\Delta_7$ clearly tracks that of the energy separation of the split peaks at the $L_3$-edge.

 With this excellent agreement between the experimental spectra and the calculation, we can speculate on a possible physical origin of the $n-$dependence of the correlated gap energy. The increasing magnitude of  $\Delta_7$ with decreasing number of LNO unit cells can be connected to the effect of the confinement which breaks the translation symmetry with interfaces and causes the formation of  a significantly  less covalent Ni-O-Al chemical bond. The  absence of $d$-electron states on Al and its much lower electronegativity compared to Ni strongly suppresses the ligand hole density on the apical oxygen O$_{a}$ \cite{Han,Freeland}. Consequently, because of the proximity to the more positively charged Al ion, the Madelung potential on the interfacial O$_{a}$ is further raised with respect to the planar O$_{p}$; this leads to further suppression on the apical hole density and charge transfer process. As a result, the interfacial Ni-O$_{a}$  bond length becomes shorter \cite{Chaloupka}, the ligand hole is transferred into the NiO$_{2}$ plane and the $d^8{\underline L}$ configuration will acquire more of the $d_{x^{2}-y^{2}}$ character. This effect is in close analogy to the physics of cuprates where the Madelung energy of the apical oxygen and the Cu-O$_{a}$ bond length are  believed to be important parameters to control the planar hole density and to stabilize the Zhang-Rice states upon hole-doping \cite{Mori}.

%\section{DMFT}

As discussed before the cluster calculations can be successfully applied to provide the insight how the local electronic structure is modified in proximity to the interface. However, the observed global MIT resulting from the competition between the short-range electron correlation and the reduced bandwidth subject to quantum confinement is better treated within the layer extension
of cellular dynamical-mean-field theory (layer CDMFT) \cite{Kotliar} with the Lanczos exact diagonalization impurity solver \cite{Caffarel}. Note that CDMFT was recently applied to explain the opening of a correlation-induced pseudogap in  SrVO$_3$ thin films \cite{Yoshimatsu}.
The layer CDMFT we employed here goes beyond that previous study by including short-range correlations within each monolayer to capture the physics of pseudogap.
To  make the problem tractable and yet relevant  for the SLs we consider the multi-layered single-band Hubbard model for Ni $3d$ and simulate the high-energy Al $3s$ states as vacuum. The short-range correlation is antiferromagnetic as recently suggested in a similar system with the planar $d_{x^{2}-y^{2}}$ orbital configuration \cite{Hansmann}. To neutralize the CDMFT's overemphasis on the in-plane correlations,
we introduce a small anisotropy between the in-plane $t_{in}$ and out-of-plane $t_{out}$ transfers as $t_{in} = t (1 - \delta)$ and $t_{out} = t(1 + 2 \delta)$  so that $t_{out}$ competes with the in-plane correlations. The bandwidth of the bulk non-interacting system is unchanged, $W=12 t$.

\begin{figure}[t]\vspace{-0pt}
\includegraphics[width=8.5cm]{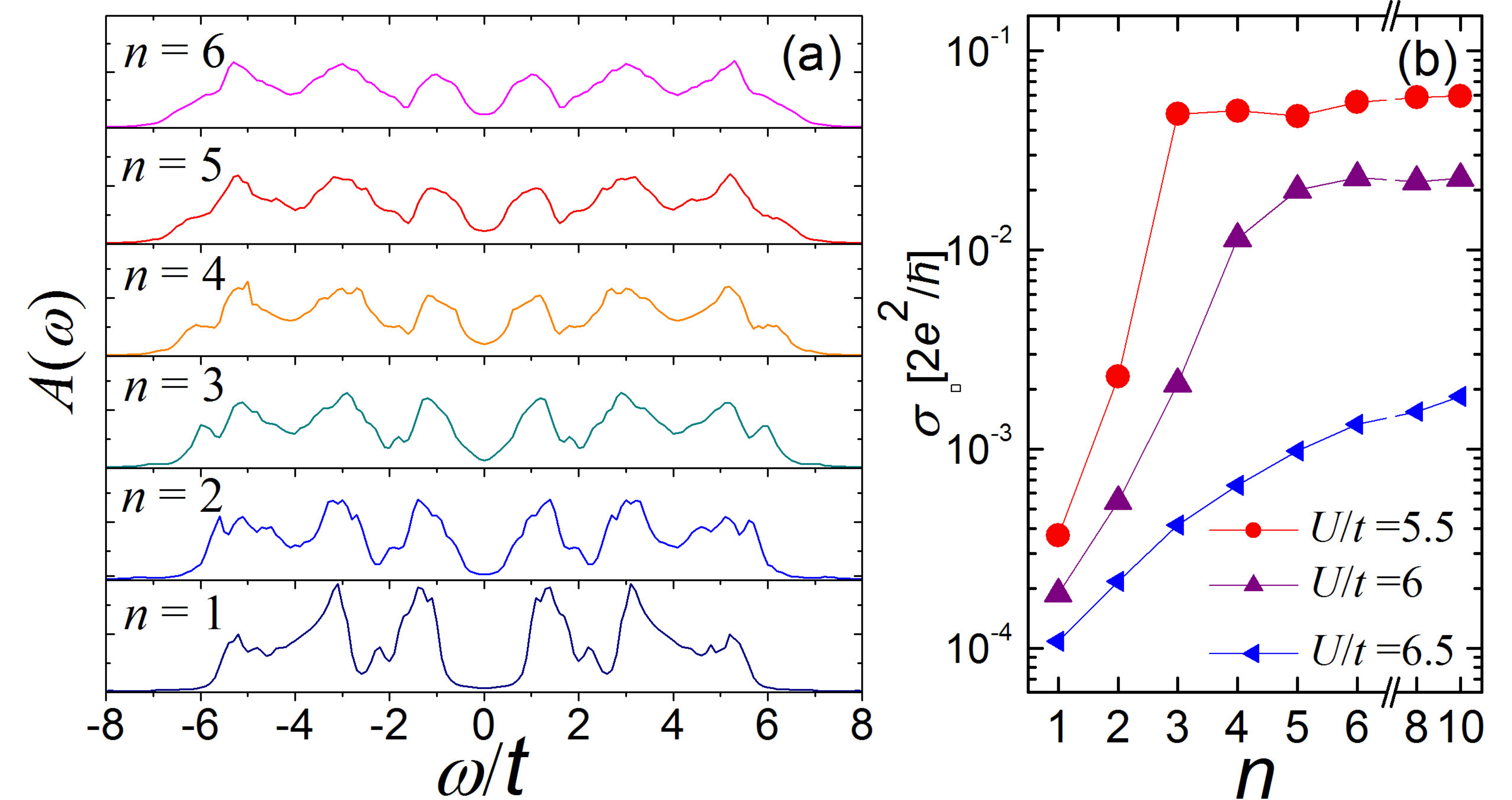}
\caption{\label{fig:theory} Layer CDMFT results for the multilayer Hubbard model with various thickness $n$. The
transfer anisotropy $\delta$ is set at 0.2.
(a) Average spectral function for $U/t = 6$.
(b) Sheet conductance $\sigma_{\Box}$ versus $n$ for $U/t = 5.5$, 6 and 6.5.
}
\end{figure}

Since bulk LNO is a metal we consider moderate values of $U/t$ ($U=6t$).
Fig.~\ref{fig:theory} (a) presents the $n$-dependent evolution of the spectral function.
As clearly  seen, the  correlation gap in the two dimensional limit is filled up and becomes a pseudogap with increasing $n$, lending theoretical  support to the notion that the experimentally observed  MIT arises from  the dimensional evolution of a quantum confined correlated carriers transforming from an insulator to a metal through a `bad' metal phase.
Notice,  the sharp quasiparticle peak is absent at $\omega =0$ even in the thickest slab because
out-of-plane correlations are treated on a mean-field level.
To see the relation between the (pseudo) gap opening and the anomalous transport properties, sheet conductance $\sigma_\Box$ is computed by the standard Kubo formula.
We introduce an imaginary part $\eta=0.1t$ to the real frequency $\omega$ to represent the finite impurity scattering.
As shown in Fig.~\ref{fig:theory} (b), with increasing $n$, $\sigma_\Box$ increases continuously and becomes practically unchanged above a critical thickness $n_c$ below which the pseudogap is pronounced.
These results reproduce well the critical regime at $n_c \approx 5$ in the transport measurements and strongly
suggest the confinement-controlled short-range electron correlations as the origin of the observed MIT. As discussed earlier, it is known that bulk rare-earth nickelates have an intrinsic propensity towards the charge disproportionation when the electronic bandwidth is reduced \cite{Catalan0}. The result is a complicated three-dimensional ordering of both charge and spin. While such an order might be thought as a complex competing state, we believe that the essential physics of  metal-insulator crossover by the quantum confinement is captured by the CDMFT.

In conclusion, we have investigated the quantum confinement effects on the electronic properties of (LNO)$_n$/(LAO)$_3$ SLs. By lowering the dimensionality of the heterostructured LNO slab, enhanced electron correlations are observed to drive the emergent Mott-type MIT with the latent competing state of charge-ordering. The strong multiplet splitting appearing in X-ray absorption spectra is the effect of the interface  leading to the redistribution of ligand hole density and reduction of  Ni-O-Al covalency as confirmed by $ab$ $initio$ cluster calculations. The result indicates intriguing similarities between the physics of heterostructured nickelate-based $e^{1}_{g}$-system and parent cuprates. The critical region of the MIT at $n=5$ deduced from the transport measurements is ascribed to the pseudogap opening  due to dimensional crossover and strongly enhanced short-range correlations confirmed by the layer-resolved CDMFT calculations.

The authors acknowledge fruitful discussions with R. Pentcheva, D. Khomskii, A. Millis and G. A. Sawatzky. J.C. was supported by DOD-ARO under the grant No. 0402-17291 and NSF grant No. DMR-0747808. Work at the Advanced Photon Source, Argonne is supported by the U.S. Department of Energy, Office of Science under grant No. DEAC02-06CH11357. MvV was  supported by the U.S. Department of Energy, Office of Basic Energy Sciences, Division of Materials Sciences and Engineering under contract DE-FG02-03ER46097.
S.O. was supported by the Materials Sciences and Engineering Division, Office of Basic Energy Sciences, the US DOE.


\begin{thebibliography}{99}
\bibitem{Bednorz}
J. G. Bednorz, Nature Mater. \textbf{6}, 821 (2007).

\bibitem{Reyren}
N. Reyren et al., Science \textbf{317}, 1196 (2007).

\bibitem{Takahashi}
H. Takahashi et al., Nature \textbf{453}, 376 (2008).

\bibitem{Chaloupka}
J. Chaloupka and G. Khaliullin, Phys. Rev. Lett. \textbf{100}, 016404 (2008); P. Hansmann et al., \textit{ibid.} \textbf{103}, 016401 (2009).

\bibitem{Hansmann}
P. Hansmann et al., Phys. Rev. B \textbf{82}, 235123 (2010).

\bibitem{Mazin}
I. I. Mazin et al., Phys. Rev. Lett. \textbf{98}, 176406 (2007).

\bibitem{Catalan0}
G. Catalan, Phase Transitions \textbf{81}, 729 (2008).

\bibitem{Han}
M. J. Han, C. A. Marianetti, and A. J. Millis, Phys. Rev. B \textbf{82}, 134408 (2010).

\bibitem{Freeland}
J. W. Freeland
et al.%, Jian Liu, M. Kareev, B. Gray, J.W. Kim, P. Ryan, R. Pentcheva and J. Chakhalian
, arXiv:1008.1518v1.

\bibitem{Nakagawa}
N. Nakagawa, H. Y. Hwang, and D. A. Muller, Nature Mater. \textbf{5}, 204 (2006).

\bibitem{Liu}
Jian Liu
et al.%, M. Kareev, S. Prosandeev, B. Gray, P. Ryan, J. W. Freeland and J. Chakhalian
, Appl. Phys. Lett. \textbf{96}, 133111 (2010).

\bibitem{May}
S. J. May et al., Phys. Rev. B \textbf{82}, 014110 (2010).

\bibitem{Chakhalian2}
J. Chakhalian
et al.%, J.M. Rondinelli, Jian Liu, B. Gray, M. Kareev, E.J. Moon, M. Varela, S.G. Altendorf, F. Strigari, B. Dabrowski, L.H. Tjeng, P.J. Ryan, J.W. Freeland
, arXiv:1008.1373v1.

\bibitem{Seo}
S. S. A. Seo
et al.%, M. J. Han, G. W. J. Hassink, W. S. Choi, S. J. Moon, J. S. Kim, T. Susaki, Y. S. Lee, J. Yu, C. Bernhard, H. Y. Hwang, G. Rijnders, D. H. A. Blank, B. Keimer, and T. W. Noh
, Phys. Rev. Lett. \textbf{104}, 036401 (2010).

\bibitem{Hotta}
M. Takizawa
et al.
, Phys. Rev. Lett. \textbf{102}, 236401 (2009).

\bibitem{Kareev1}
M. Kareev
et al.%, S. Prosandeev, Jian Liu, B. Gray, P. Ryan and J. Chakhalian
, arXiv:1005.0570v1.

\bibitem{Kareev2}
M. Kareev
et al.%, S. Prosandeev, J. Liu, C. Gan, A. Kareev, J. W. Freeland, Min Xiao and J. Chakhalian
, Appl. Phys. Lett. \textbf{93}, 061909 (2008).


\bibitem{Zhou}
J.-S. Zhou et al., Phys. Rev. B \textbf{61}, 4401 (2000); K. Sreedhar et al., \textit{ibid.} \textbf{46}, 6382 (1992).

\bibitem{Rivadulla}
F. Rivadulla, J.-S. Zhou, and J. B. Goodenough, Phys. Rev. B \textbf{67}, 165110 (2003).


\bibitem{Scherwitzl}
R. Scherwitzl
et al.%, P. Zubko, C. Lichtensteiger, and J.-M. Triscone
, Appl. Phys. Lett. \textbf{95}, 222114 (2009); J. Son et al., \textit{ibid.} \textbf{96}, 062114 (2010).

\bibitem{Mott}
N. F. Mott, Philos. Mag. \textbf{26}, 1015 (1972).


\bibitem{footnote1}
The partially overlapping La $M_4$-edge is substracted from the Ni $L_3$-edge spectra.

\bibitem{Freeland2}
J. W. Freeland et al., unpublished.

\bibitem{Medarde}
M. Medarde
et al.%, A. Fontaine, J.L. Garcia-Mu?oz, J. Rodriguez-Carvajal, M. de Santis, M. Sacchi, G. Rossi, and P. Lacorre
, Phys. Rev. B \textbf{46}, 14975 (1992).

\bibitem{Piamonteze}
C. Piamonteze
et al.%, F. M. F. de Groot, H. C. N. Tolentino, A. Y. Ramos, N. E. Massa J. A. Alonso and M. J. Martinez-Lope
, Phys. Rev. B \textbf{71}, 020406(R) (2005).

\bibitem{Veenendaal}
M. A. van Veenendaal and G. A. Sawatzky, Phys. Rev. B \textbf{50}, 11326 (1994).

\bibitem{Mori}
M. Mori et al., Phys. Rev. Lett. \textbf{101}, 247003 (2008).

\bibitem{Kotliar}
G. Kotliar
et al.%, S. Y. Savrasov, G. P{\'a}lsson, and G. Biroli
, Phys. Rev. Lett. \textbf{87}, 186401 (2001).

\bibitem{Yoshimatsu}
K. Yoshimatsu et al., Phys. Rev. Lett. \textbf{104}, 147601 (2010).

\bibitem{Caffarel} M. Caffarel and W. Krauth, Phys. Rev. Lett. \textbf{72}, 1545 (1994).



\end{thebibliography}
\end{document}